\documentclass[11pt,a4paper]{article}
\usepackage{jheppub}
\usepackage[utf8]{inputenc}
\usepackage{silence}
\usepackage{bm}
\usepackage{longtable}
\usepackage[missing=]{gitinfo2}
\usepackage{xcolor}

\preprint{{\flushright
UTTG--02--19\\
}}
\title{BPS states in the Minahan-Nemeschansky $E_7$ theory}
\author[1]{Qianyu Hao,}
\author[2]{Lotte Hollands,}
\author[3]{Andrew Neitzke}
\affiliation[1]{Department of Physics, University of Texas at Austin}
\affiliation[2]{Department of Mathematics, Heriot-Watt University}
\affiliation[3]{Department of Mathematics, University of Texas at Austin}
\abstract{We use the method of spectral networks to calculate BPS degeneracies 
in the Minahan-Nemeschansky $E_7$ theory, 
as representations of the $E_7$ flavor symmetry.
Our results provide another example of a pattern noticed earlier in 
the Minahan-Nemeschansky $E_6$ theory:
when the electromagnetic charge is $n$ times a primitive charge,
the BPS index is a positive integer multiple of $(-1)^{n+1} n$. 
We also calculate BPS degeneracies in the Minahan-Nemeschansky $E_6$
theory for larger charges than were previously computed.}
\arxivnumber{}   
\usepackage{blindtext}
\usepackage{amsmath}
\usepackage{upgreek}
\usepackage{amssymb}
\usepackage{slashed}
\usepackage{mathtools}
\usepackage{mleftright}
\usepackage{adjustbox}

\newcommand{\CP}{{\mathbb {CP}}}
\newcommand{\cN}{{\mathcal N}}
\newcommand{\de}{{\mathrm d}}
\newcommand{\e}{{\mathrm e}}
\newcommand{\I}{{\mathrm i}}
\newcommand{\Z}{{\mathbb Z}}
\newcommand{\red}{{\mathrm{red}}}

\makeatletter
\newcommand{\tpitchfork}{%
  \vbox{
    \baselineskip\z@skip
    \lineskip-.52ex
    \lineskiplimit\maxdimen
    \m@th
    \ialign{##\crcr\hidewidth\smash{$-$}\hidewidth\crcr$\pitchfork$\crcr}
  }%
}
\makeatother
\usepackage[utf8]{inputenc}
\begin{document}
{{{\tiny \color{gray} \tt \gitAuthorIsoDate}}
{{\tiny \color{gray} \tt \gitAbbrevHash}}}
\maketitle
\flushbottom
\section{Introduction}
\label{sec:intro}

Minahan and Nemeschansky discovered $\mathcal{N}=2$ superconformal theories in four dimensions with flavor symmetry $E_6$, $E_7$ and $E_8$~\cite{Minahan:1996fg,Minahan:1996cj}. These remarkable 
theories have been studied extensively since then. 
In this paper we study the BPS spectrum of the $E_7$ theory.

We consider the class $S$ construction~\cite{Gaiotto:2009we,Gaiotto:2009hg}, 
applied with 
Lie algebra $A_3$ and Riemann surface $C=\mathbb{CP}^1\setminus\{z_1,z_2,z_3\}$, where $z_1$, $z_2$ are full punctures
and $z_3$ is a puncture of type $[2,2]$.
This construction produces a superconformal $\cN=2$ theory with manifest flavor symmetry $SU(4)\times SU(4)\times SU(2)$.\footnote{$E_7$ does not have a subgroup isomorphic to $SU(4)\times SU(4)\times SU(2)$, but has subgroup isomorphic to $SU(4)\times SU(4)\times SU(2)/\mathbb{Z}_4$. The $\mathbb{Z}_4$ is the subgroup $\{(1,1,1)$, $(\mu,\mu,\mu^2)$, $(\mu^2,\mu^2,1)$, $(\mu^3,\mu^3,\mu^2)\}$, where $\mu$ is a primitive fourth root of unity. Thus the enhancement of symmetry requires that this $\mathbb{Z}_4$ acts trivially.} 
 Already in \cite{Gaiotto:2009we}, Gaiotto proposed that in this class $S$ 
theory the manifest $SU(4)\times SU(4)\times SU(2)$ should actually be enhanced to
an $E_7$ flavor symmetry, and indeed the theory should be the Minahan-Nemeschansky $E_7$ theory.
Some checks of this proposal are given by Benini, Benvenuti and Tachikawa in~\cite{Benini:2009gi}
and by Tachikawa in~\cite{Tachikawa:2013kta}.

Having this class $S$ realization of the Minahan-Nemeschansky $E_7$
theory allows us to study its BPS states using the method of spectral networks
\cite{Gaiotto:2012rg}, and this is what we do in this paper.
Our approach is mostly parallel to what was done for the $E_6$ 
theory in \cite{Hollands:2016kgm}, and thus we are rather brief,
focusing mainly on those points which are different for the $E_7$
theory;
see \cite{Hollands:2016kgm} for more background and explanations
of the method.

\medskip
On the Coulomb branch, the Hilbert space is graded by electromagnetic charge corresponding to the $U(1)$ gauge symmetry. The electromagnetic charge lattice has rank $2$, and can be identified with the homology $H_1(\overline\Sigma, \Z)$ where $\overline\Sigma$ is the Seiberg-Witten curve, given below as \eqref{eq:sw-explicit}. 
We introduce a basis $\{\gamma_1, \gamma_2\}$ for
this charge lattice, where $\langle\gamma_1,\gamma_2\rangle=1$;
we call $\gamma_1$ the primitive
electric charge, and $\gamma_2$ the primitive magnetic charge.
A general charge can be written as $\gamma=p \gamma_1 + q \gamma_2$.
The theory has a $\mathbb{Z}_2$ symmetry, induced from the symmetry of $C$ which exchanges the two full punctures; this symmetry swaps the charges 
$\gamma_1 \leftrightarrow \gamma_2$.

The main new result in this paper is the computation of BPS indices for various charges,
of the form $n \gamma_1$ and $n(\gamma_1 + \gamma_2)$, as we now describe.

The spectral network relevant for computing BPS indices for particles 
with charges $n \gamma_1$ looks like a circle; it is shown in \autoref{fig:ja}
below. 
Using this spectral network we have computed the indexed counts 
$\Omega(n\gamma_1)$ of 4d BPS states for $1 \le n \le 200$;
for $1 \le n \le 11$ the results are given in \autoref{table:numerical-indices}.
For example, we find 
\begin{equation}
\Omega(9\gamma_1)= 292459392000.
\end{equation}
We also show that the BPS index has asymptotic exponential growth
\begin{equation}
|\Omega(n\gamma_1)| \sim cn^{-\frac{5}{2}}(17 + 12 \sqrt{2})^{n}.
\end{equation}
Our computation has manifest $SU(4) \times SU(4) \times SU(2)$ flavor symmetry, so the integers
$\Omega(\gamma)$ admit an ``upgrade'' to characters $\mathbf{\Omega}(\gamma)$ of representations of $SU(4) \times SU(4) \times SU(2)$.
Since the flavor symmetry is predicted to be enhanced to $E_7$,
these characters should \textit{a posteriori} assemble into 
characters of representations of $E_7$.
We compute $\mathbf{\Omega}(n\gamma_1)$ for $1\leq n\leq 11$, 
and find that indeed they are characters of $E_7$, as predicted. For example, we find 
\begin{equation} \label{eq:flavored-example-1}
\mathbf{\Omega}(3\gamma_1) = 3 \times \bm{912}+ 6 \times\bm{56}.
\end{equation}
The full results are given in \autoref{table:electric-indices} below.

We also calculate the BPS index for the charges $n(\gamma_1+\gamma_2)$ with $1 \le n \le 5$;
this involves a different spectral network, shown in \autoref{fig:charge11}.
We obtain, for example,
\begin{equation} \label{eq:flavored-example-2}
\bm{\Omega}(2(\gamma_1+\gamma_2))=- 2 \times\bm{1539} - 4\times\bm{133} - 8\times\bm{1}.
\end{equation}
The full results are in \autoref{table:11-indices} below.

Our results exhibit the same pattern observed in the $E_6$ Minahan-Nemeschansky theory \cite{Hollands:2016kgm}: 
BPS states carrying electromagnetic charges which are $n$ times a primitive charge always
come with index a positive integer multiple of $(-1)^{n+1} n$. For example, in \eqref{eq:flavored-example-1} 
above, the multiplicities of
irreducible representations of $E_7$ are $3$ and $6$, which are positive integer multiples of $3$;
similarly in \eqref{eq:flavored-example-2} above, all the multiplicities are positive integer
multiples of $-2$.

BPS states for rank 1 Minahan-Nemeschansky theories 
were recently studied in~\cite{Distler:2019eky}, using string junctions
in the $F$-theory realization. In that paper the precise BPS multiplicities
were not computed; rather, what was computed is a list of which
representations of the flavor symmetry group can occur in the spectrum, 
for each electromagnetic charge.
Our results for the $E_7$ theory are all in agreement with the lists
of representations given in~\cite{Distler:2019eky}.

Finally, we revisit the BPS states of the Minahan-Nemeschansky $E_6$ theory,
extending the results of~\cite{Hollands:2016kgm} to higher charges: see
\autoref{table:electric-indices-e6} below.

\section*{Acknowledgements}

We thank Jacques Distler and Mario Martone for helpful discussions.
LH's work on this paper is supported by a Royal Society Dorothy Hodgkin
Fellowship.
QH's and AN's work on this paper is supported by NSF grant DMS-1711692.

\section{Seiberg-Witten curve} 

The IR $U(1)$ gauge theory of the theory 
on its Coulomb branch is described by the Seiberg-Witten curve,
which takes the form
\begin{equation} \label{eq:sw-curve-1}
\text{det}(\lambda-\Phi(z))=0,
\end{equation}
where $\Phi(z)$ is the Higgs field in the corresponding Hitchin system.
\eqref{eq:sw-curve-1} can also be written as
\begin{equation} \label{eq:sw-with-diffs}
\lambda^4+\phi_2(z)\lambda^{2}+\phi_3(z)\lambda+\phi_4(z)=0, 
\end{equation}
where $\phi_{d}(z)$ are meromorphic differentials on $C$, degree-$d$ invariant polynomial combinations of the eigenvalues of $\Phi(z)$. 
As discussed in \cite{Gaiotto:2009we}, and reviewed 
e.g. in~\cite{Chacaltana:2012zy}, the form of $\Phi(z)$ 
around a puncture is $\Phi(z) = (\frac{\Phi_{-1}}{z}+\Phi_{0}+\dots) \, \de z$, where $\Phi_{-1}$ is constrained
to lie in a specific nilpotent orbit of $\mathfrak{sl}_4$, determined by a partition $\rho$ of $4$. 
$\rho$ determines the Jordan block structure of $\Phi_{-1}$, which in turn
determines the form of the meromorphic differentials $\phi_{d}(z)$;
$\phi_{d}(z)$ has a pole of order at most $p_{d}(\rho)$ 
at a puncture with partition $\rho$. 

In our case the two full punctures have $\rho = [4]$ and for the third puncture $\rho = [2,2]$;
the two full punctures have $(p_2,p_3,p_4) = (1,2,3)$ and the other puncture has $(p_2,p_3,p_4) = (1,1,2)$.

Using these constraints, the only nonzero differential in the Minahan-Nemeschansky $E_7$ theory turns out to be $\phi_4$.
Using the $PSL(2,\mathbb{C})$ symmetry of $\CP^1$, the three punctures
can be fixed to $(z_1 = 1, z_2 = \omega, z_3 = \omega^2)$, where $\omega = e^{2\pi i/3}$. Then the differentials are concretely
\begin{equation}
\phi_2=0, \qquad \phi_3=0, \qquad \phi_4=-\frac{u\,\de z^4}{(z-1)^3(z-\omega)^3(z-\omega^2)^2}.
\end{equation}
The free parameter $u\in \mathbb{C}$ parameterizes the 1-complex-dimensional Coulomb branch.
By the scale invariance and $U(1)_R$ symmetry of the theory, 
all points with $u \neq 0$ are equivalent, so from now on we set $u=1$.
We also write $\lambda = x \, \de z$. 
Then the Seiberg-Witten curve \eqref{eq:sw-with-diffs} becomes
\begin{equation} \label{eq:sw-explicit}
\Sigma=\left\{x^4-\frac{1}{(z-1)^3(z-\omega)^3(z-\omega^2)^2}=0\right\}.
\end{equation}
By filling in the punctures, we get a smooth compact genus 1 curve $\overline{\Sigma}$.
The projection $\pi:\overline\Sigma\rightarrow C$ is a degree $4$ covering,
branched over the $3$ punctures on $C$.

\begin{figure}[tbp]
\centering 
\includegraphics[width=.45\textwidth]{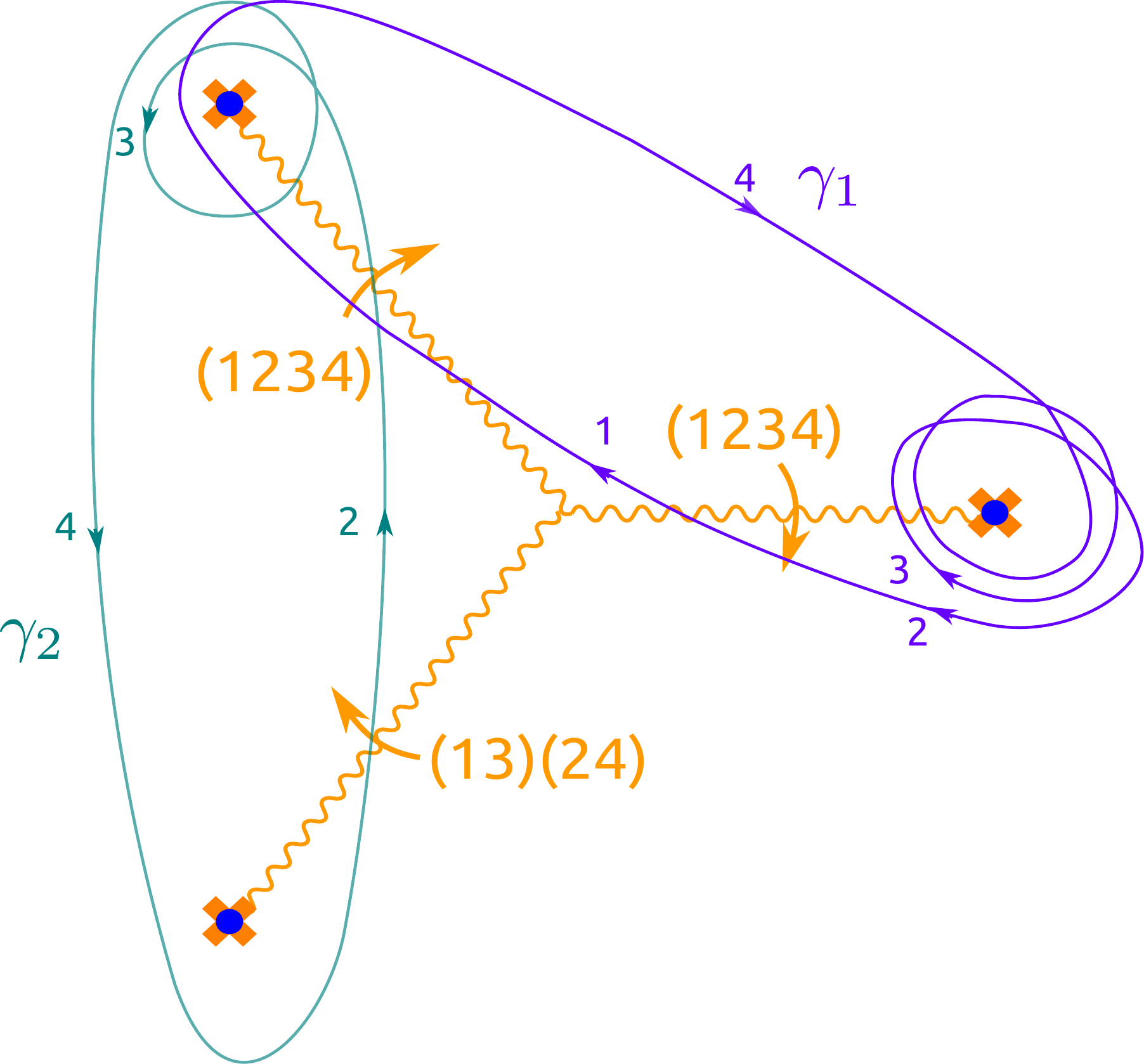}
\caption{\label{fig:i} Homology classes on $\overline{\Sigma}$ representing the primitive electric charge $\gamma_1$ and primitive magnetic charge $\gamma_2$. The purple cycle is the primitive electric charge $\gamma_1$ and the green cycle is the primitive magnetic charge $\gamma_2$. The intersection number $\langle\gamma_1,\gamma_2\rangle$ is 1. The wavy lines carrying permutations are the branch cuts in our presentation of $\Sigma$ as a $4$-fold covering of the plane. The three punctures are at $z_1 = 1$, $z_2 = \omega$ and $z_3 = \omega^2$. The numbers on each path segment indicate which sheet the path is on.}
\end{figure}

The electromagnetic charge lattice of the IR $U(1)$ theory on the Coulomb branch is
\begin{equation}
\Gamma_g=H_1(\overline{\Sigma},\mathbb{Z}).
\end{equation}
Two basis charges $\gamma_1$ and $\gamma_2$
are sketched in \autoref{fig:i}; by convention we call $\gamma_1$ ``electric''
and $\gamma_2$ ``magnetic.'' The central charge corresponding to the EM charge $\gamma$ is given by the integral $Z_{\gamma}=\frac{1}{\pi}\oint_{\gamma}\lambda$.
For the primitive electric charge this gives
\begin{equation}Z_{\gamma_1}=\frac{1}{\pi}\int_1^{\omega}\lambda_1+\frac{1}{\pi}\int^1_{\omega}\lambda_4=\frac{4\sqrt{\frac{2}{3}}}{\pi^{3/2}}\Gamma\left[\frac{5}{4}\right]\Gamma\left[\frac{1}{4}\right] \e^{-\frac{7 \pi}{12}\I}\approx 1.92749 \e^{-\frac{7 \pi}{12}\I},
\end{equation}
and for the primitive magnetic charge
\begin{equation}
Z_{\gamma_2} = -\I Z_{\gamma_1}=\frac{1}{\pi}\int_\omega^{\omega^2}\lambda_4+\frac{1}{\pi}\int^{\omega}_{\omega^2}\lambda_2=\frac{4\sqrt{\frac{2}{3}}}{\pi^{3/2}}\Gamma\left[\frac{5}{4}\right]\Gamma\left[\frac{1}{4}\right]\e^{\frac{11 \pi}{12}\I}\approx 1.92749 \e^{\frac{11 \pi}{12}\I}.
\end{equation}

Notice that the curve $\Sigma$ given in \eqref{eq:sw-explicit} has $\mathbb{Z}_4$ symmetry,
generated by the transformation $\lambda \mapsto \I \lambda$. This generator permutes the primitive
electric and magnetic charges: it maps $\gamma_1 \mapsto -\gamma_2$ and
 $\gamma_2 \mapsto \gamma_1$. 

\section{Computing the BPS states}
\label{sec:BPS states}

We use the same spectral network technique for computing BPS states as was used for the Minahan-Nemeschansky $E_6$ theory in \cite{Hollands:2016kgm}. The surface $C$ is a parameter space for surface defects; the spectral network $W(\vartheta)$ consists of the points $z$ such that the surface defect with parameter $z$ supports a BPS soliton of central charge $Z$ with phase $\arg(-Z)=\vartheta$. To study the bulk BPS states of charge $\gamma$, we must choose the phase of the spectral network to be $\vartheta = \vartheta_{\gamma}=\arg(-Z_{\gamma})$.

\subsection{Building the spectral network} \label{sec:building-sn}

We comment briefly on how we compute the spectral network $W(\vartheta)$. 
The network is made up of ``$S$-walls'' which obey
differential equations.
In addition to its parameterization $z(t)$, each wall carries a pair of labels $ij$ and 
two auxiliary functions $x_1(t)$ and $x_2(t)$: if the wall is labeled $ij$ then $x_1(t)$
gives the $i$-th sheet of $\Sigma$ over the wall, and $x_2(t)$ the $j$-th sheet.
The condition $\arg(-Z) = \vartheta$ translates into differential equations
which control the 
$t$ dependence of each wall:\footnote{Alternatively one could say that
only $z(t)$ is determined by a differential equation, while
$x_1(z)$, $x_2(z)$ are just solutions of the algebraic equation 
$F(x,z(t)) = 0$ which move continuously with $t$.
This continuity condition is hard to implement in practice, because of the branch cut of the 
fourth-root function. The
advantage of writing differential equations for $x_1(t)$, $x_2(t)$ as well as for $z(t)$ 
is that it automatically enforces the continuity, thus avoiding having to deal with cuts. This method is implemented in \cite{swn-plotter}.}
\begin{equation} \label{eq:s-wall-eq-1}
z'(t) = -(x_1(t)-x_2(t))^{-1} \e^{\I \vartheta}, \quad x_1'(t)=\frac{\de x}{\de z} z'(t), \quad x_2'(t)=\frac{\de x}{\de z} z'(t),
\end{equation}
where
\begin{equation} \label{eq:s-wall-eq-2}
\frac{\de x}{\de z}=\frac{\frac{\partial F}{\partial z}}{\frac{\partial F}{\partial x}}, \quad F(x,z) = x^4 + \phi_4(z).
\end{equation}

\begin{figure}[tbp]
\centering 
\includegraphics[width=.35\textwidth]{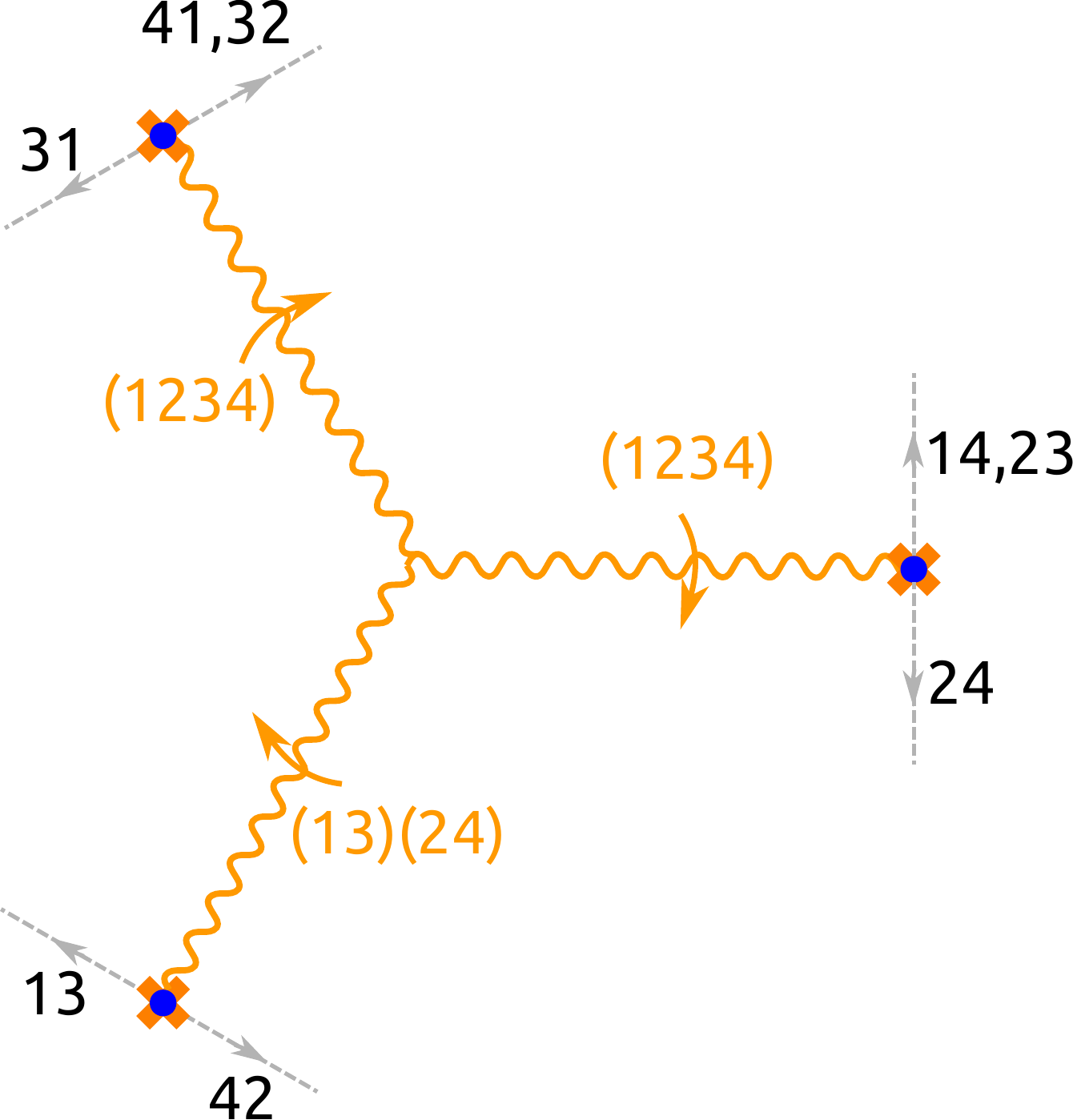}
\caption{\label{fig:snstubs}{The distinguished directions and sheet labels
emerging from the punctures, at $\vartheta = \vartheta_{\gamma_1} = \frac{5\pi}{12}$. 
These serve as seeds
for the $S$-walls making up the spectral network.}}
\end{figure}

Next we need to explain the initial conditions: where do new walls originate?
There are a few possibilities: either branch points of the covering $\Sigma \to C$,
or from places where existing walls intersect one another. 
In the example we consider
here, we will only have to deal with the case of walls originating from branch points.
In this theory the branch points coincide with the punctures.
Thus we need to study solutions of \eqref{eq:s-wall-eq-1} originating at a puncture.
Integrating \eqref{eq:s-wall-eq-1} 
from the puncture $z_i$ to some nearby point
$z_i + \delta$ gives the constraint 
$-(\e^{-\I\vartheta}\int_{z_i}^{z_i+\delta}(x_1(z)-x_2(z))\, \de z)\in\mathbb{R}_{+}$.
For small $\delta$, and a wall of type $ij$,
this integral is proportional to $(\e^{\I \pi i/2} - \e^{\I \pi j/2}) \delta^{\frac14}$
for a full puncture, or to $(\e^{\I \pi i/2} - \e^{\I \pi j/2}) \delta^{\frac12}$ for a type $[2,2]$ puncture;
thus, in either case, this constraint singles out distinguished directions $\arg (\delta)$ 
and sheet labels $ij$. We also impose the additional constraint that 
$x_1 = x_2$ at the puncture.
See \autoref{fig:snstubs} for the resulting distinguished directions and sheet labels,
at the phase $\vartheta = \vartheta_{\gamma_1} = \frac{5\pi}{12}$.
(Note that two different walls emerging
from a puncture can be exactly degenerate: e.g. emanating from the puncture at
$z=z_1=1$ there is a wall labeled $14$ which is 
exactly degenerate with a wall labeled $23$.)

We then choose initial points $z = z_i + \delta$
very close to the punctures, and initial values $x_1, x_2$ determined by the sheet labels,
and numerically integrate the equations
\eqref{eq:s-wall-eq-1}-\eqref{eq:s-wall-eq-2} to determine the full $S$-walls.
After finite time, we find that the $S$-walls emerging from one puncture run into
a neighboring puncture; at that point we just terminate them.

We make the most conservative possible assumption, that any $S$-wall allowed by this
analysis indeed exists. (This assumption will be verified in the next section when 
we compute the soliton counts on the walls and see that they are nonzero.)

\begin{figure}[tbp]
\centering 
\includegraphics[width=.3\textwidth]{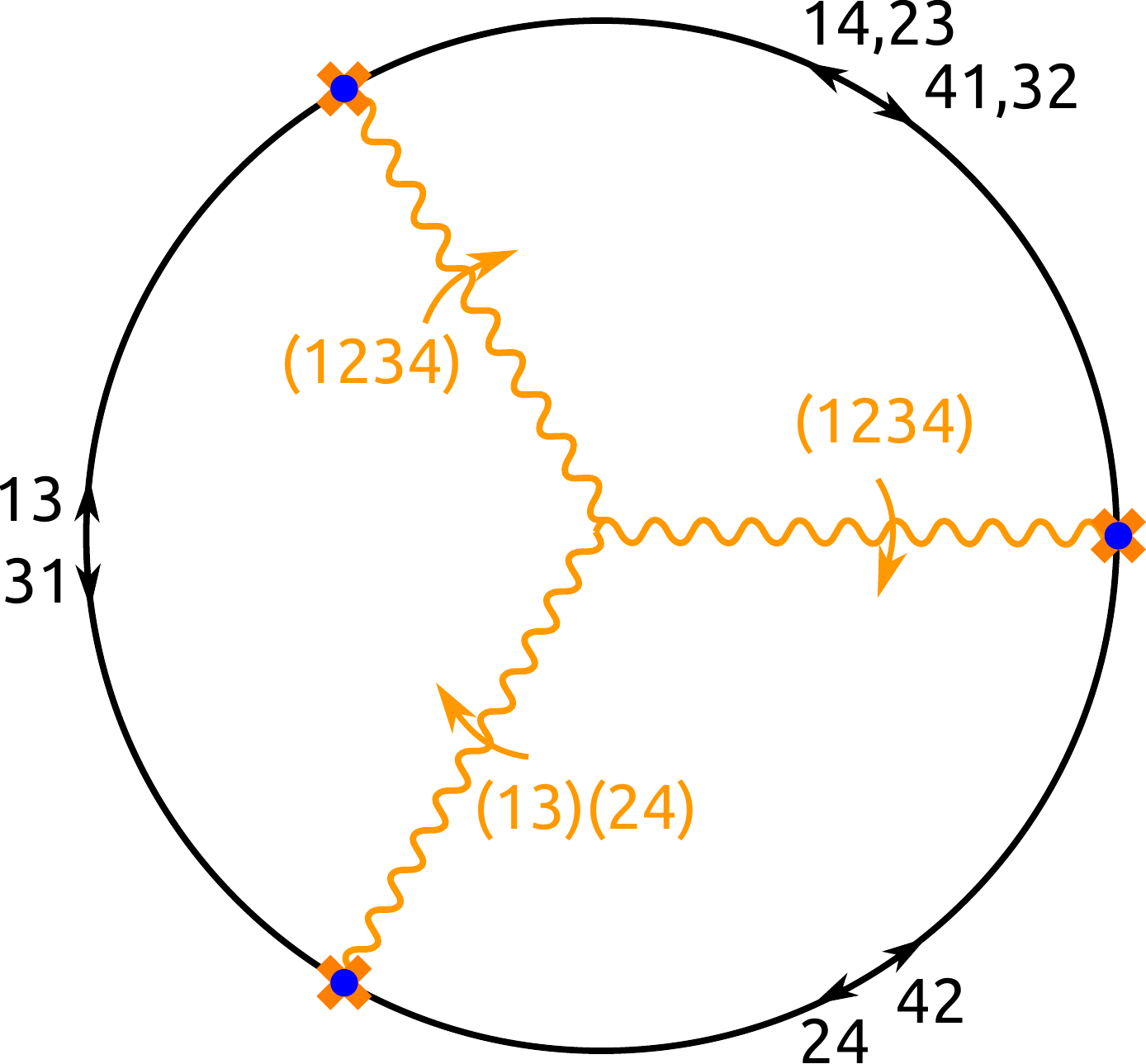}
\caption{\label{fig:ja}The spectral network of the theory at the phase $\vartheta = \frac{5\pi}{12}$;
this is the network relevant for computing BPS states with charge $n\gamma_1$.
All $3$ arcs shown support ``double walls,'' i.e. superpositions of a wall
of type $ij$ and a wall of type $ji$, with opposite orientations. The arc at the northeast
is even more degenerate: it is a superposition of two double walls, one of type
$14/41$ and one of type $23/32$.}
\end{figure}

The outcome of this process at the phase $\vartheta = \vartheta_{\gamma_1} = \frac{5\pi}{12}$ is the very simple 
spectral network shown in \autoref{fig:ja}.

\subsection{Finding the solitons}

\begin{figure}[tbp]
\centering 
\includegraphics[width=.3\textwidth]{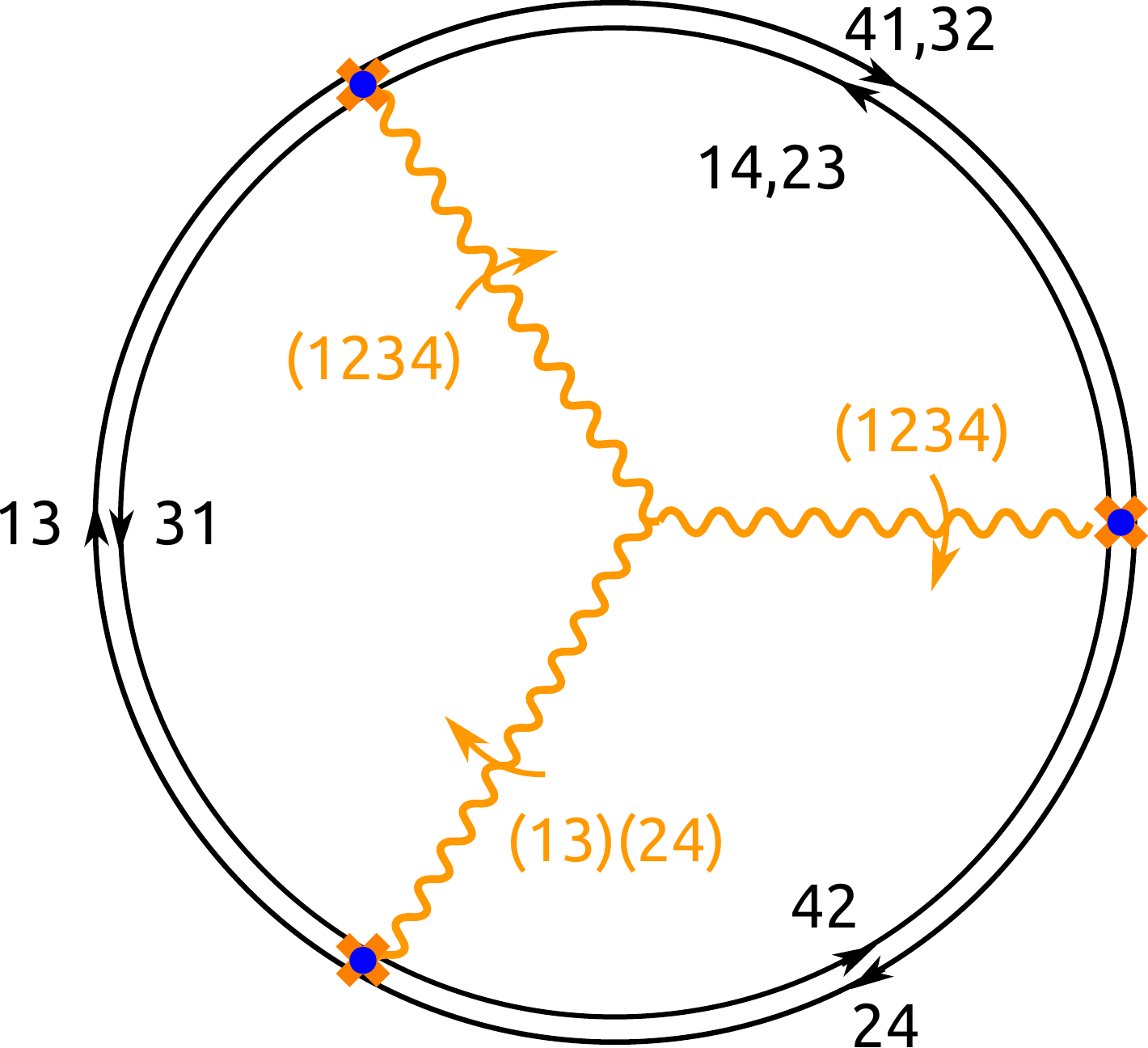}
\caption{\label{fig:resolutionofcharge1}Resolution of the spectral network of \autoref{fig:ja}. 
Each double wall from \autoref{fig:ja} has been replaced by two infinitesimally separated walls.
The walls carrying labels $14$ and $23$ are on top of one another
even after the resolution, and likewise for those with labels $41$ and $32$.}
\end{figure}

In order to determine the bulk BPS indices, following the strategy in 
\cite{Hollands:2016kgm},
we first deform $\vartheta$ infinitesimally to get a resolution of the spectral network, as shown in \autoref{fig:resolutionofcharge1}.

\begin{figure}[tbp]
\centering 
\includegraphics[width=0.95\textwidth]{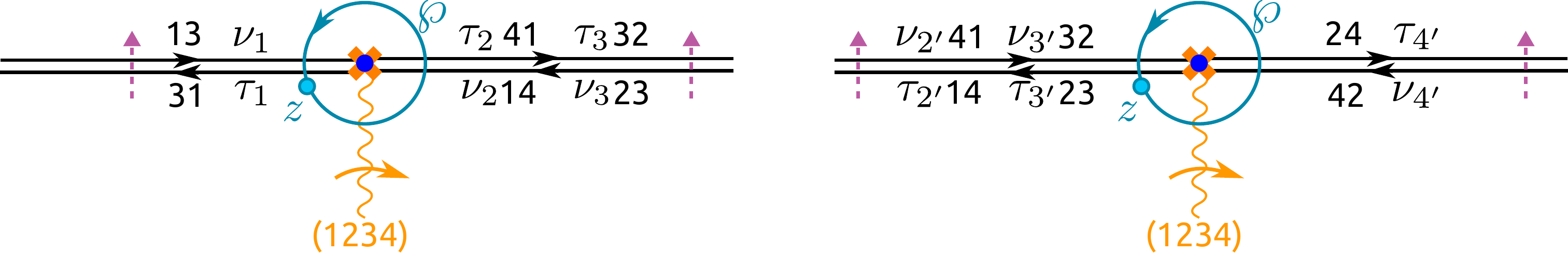}
\caption{\label{fig:loop-1}The loops around the full punctures $z_2 = \omega$ (left) and $z_1 = 1$ (right) for which we compute the framed 2d-4d BPS spectrum.}
\end{figure}

\begin{figure}[tbp]
\centering 
\includegraphics[width=.45\textwidth]{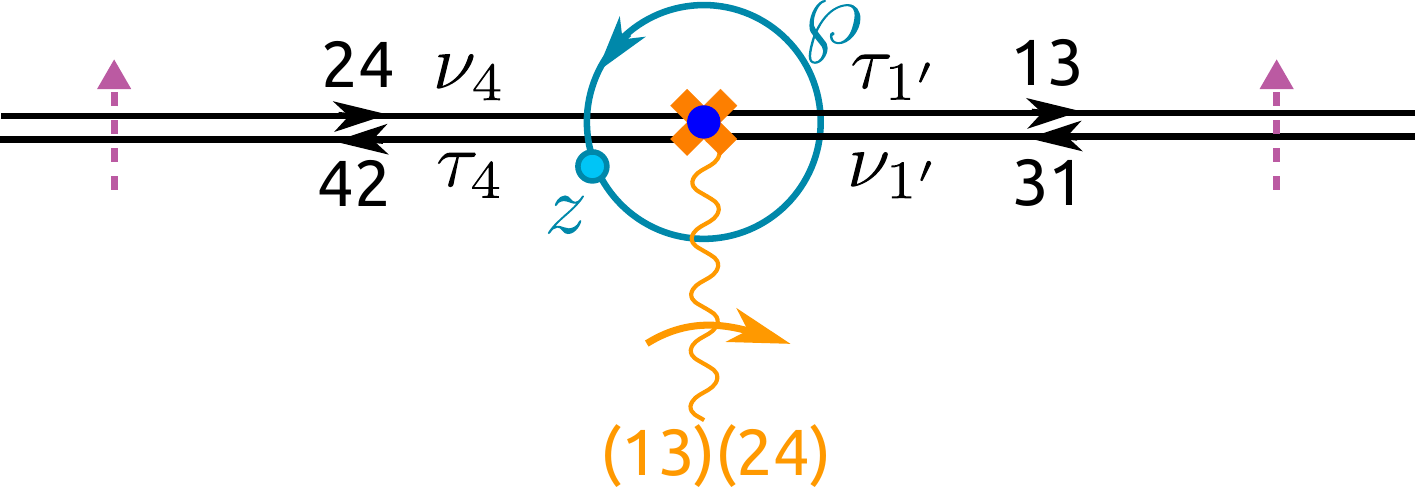}
\caption{\label{fig:loop-2}The loop around the $[2,2]$ puncture $z_3 = \omega^2$ for which we compute the framed 2d-4d BPS spectrum.}
\end{figure}

Next we apply the constraints of homotopy invariance for 2d-4d framed BPS spectra, as described in Section 4.3 of~\cite{Hollands:2016kgm}, to the resolved spectral network.
This involves studying the generating functions of 2d-4d framed BPS states associated to three
different interfaces, associated to loops around the three punctures.
See \autoref{fig:loop-1} and \autoref{fig:loop-2} for the loops we consider,
and \autoref{app:signs} for more details about the generating
functions and formal variables we use below.

The generating functions of 2d-4d framed BPS states associated to these three loops have the form:
\begin{align}
\bm{M}_1 &= \bm{F}_{p'_{1}}(1+\tau_{3'})(1+\tau_{2'})(1+\nu_{3'})(1+\nu_{2'})\bm{F}_{p_1}(1-\tau_{4'})(1-\nu_{4'}), \\
\bm{M}_2 &= \bm{F}_{p'_2}(1+\tau_{1})(1+\nu_1)\bm{F}_{p_2}(1-\tau_{3})(1-\tau_{2})(1-\nu_3)(1-\nu_2),\\
\bm{M}_3 &= \bm{F}_{p'_3}(1+\tau_{4})(1+\nu_4)\bm{F}_{p_3}(1-\tau_{1'})(1-\nu_{1'}).
\end{align}
Here in each case $p$ and $p'$ denote two semicircles on $C$, 
making up a circular loop $\wp$ around a puncture; $p$ is the top half and $p'$ the bottom half.
$p$ does not cross a branch cut, while $p'$ does cross a cut, thus going from one sheet to the next according to the permutation attached to the cut. The generating function $\bm{F}_{p}$ is the sum of 
formal variables $X_{p^{(i)}}$ associated to the four lifts of $p$ to $\Sigma$,
and similarly $\bm{F}_{p'}$. Finally, $\tau_i$ or $\nu_i$ denote the soliton generating functions.
Each of these functions counts BPS solitons with charges of the form $a + n \gamma_1$, $n \ge 0$,
where $a$ is a basic soliton charge; thus the function is of the form
$f(x) X_a$, where $X_a$ is the formal variable standing in for soliton charge $a$,
and $x = - X_{\widetilde\gamma_1}$ is the formal variable standing in for 4d particle
charge $\gamma_1$ (see \autoref{app:signs} for more details on this tricky sign.)
These functions $f(x)$ are the main undetermined quantities which we need to find.
As it happens, the spectral network in \autoref{fig:resolutionofcharge1}
contains at most one wall of any type $ij$;
thus we can distinguish the various $f(x)$ by labeling them $f_{ij}(x)$,
and our job is to determine the eight functions $f_{13}$, $f_{31}$, $f_{14}$, $f_{41}$,
$f_{23}$, $f_{32}$, $f_{42}$, $f_{24}$.

The $\bm{M}_l$, $l=1,2,3$, are naturally viewed as $4 \times 4$ matrices, 
since they contain counts of solitons going from vacuum $i$ to vacuum $j$
with $1 \le i,j \le 4$. Moreover, since the solitons are charged
under the flavor symmetry, the $\bm{M}_l$ can be promoted from numbers to characters
depending on a flavor parameter $g\in SU(4)\times SU(4) \times SU(2)$.
As in \cite{Hollands:2016kgm}, we impose the constraint that
the characteristic polynomial of $\bm{M}_l$ equals the characteristic polynomial of $g$ 
acting in a representation $R_l$. 
$R_l$ is the representation of $SU(4) \times SU(4) \times SU(2)$ which is the fundamental
for the $l$-th factor, and the trivial representation of the other two factors. 
Explicitly, for the full punctures, this characteristic polynomial can be written as
\begin{equation}
\det{
\begin{pmatrix} m_{1}-t & 0& 0&0 \\
   0 & m_{2}-t &0 &0 \\
    0& 0& m_{3}-t &0 \\
    0& 0& 0& m_4-t
    \end{pmatrix}}
=1-{\bf 4}t+{\bf 6}t^2-\bar{\bf 4}t^3+t^4,
\end{equation}
where $m_i$ are the eigenvalues of $g$ acting in $R_l$,
obeying $m_1 m_2 m_3 m_4 = 1$. 
Combinations of their products give characters of other representations of $SU(4)$, which we denote by bold numbers: e.g. ${\bf 4} = \frac{1}{m_1} + \frac{1}{m_2} + \frac{1}{m_3} + \frac{1}{m_4}$.  So, explicitly, our constraint at each full puncture is
\begin{equation} \label{eq:constraint-1}
	\det (\bm{M}_l - t) = 1-{\bf \bar 4}t+{\bf 6}t^2-{\bf 4}t^3+t^4.
\end{equation}
For the type $[2,2]$ puncture, we impose a stronger constraint:
\begin{equation} \label{eq:jordan}
\bm{M}_3^2 -{\bf 2} \bm{M}_3 + {Id}_{4 \times 4} = 0, 
\end{equation}
where $\bm{2}=m+1/m$ is the character of the fundamental representation of $SU(2)$. Since the monodromy matrix $\bm{M}_3$ has determinant $1$, and \eqref{eq:jordan} requires the diagonal elements to be either $m$ or $1/m$, it implies that the eigenvalues are $\{m,m,1/m,1/m\}$.
However, \eqref{eq:jordan}
is stronger than just fixing the eigenvalues: it also rules out 
nontrivial Jordan blocks.

Using the constraints \eqref{eq:constraint-1}, \eqref{eq:jordan} for all three
punctures simultaneously we obtain a system of equations for the $f_{ij}(x)$.
We have not found a closed solution, but making the assumption that all 
of the $f_{ij}(x)$ have series expansions in nonnegative powers of $x$,
we can solve the equations iteratively in powers of $x$.
For example, we find:
\begin{align}
	f_{14}(x) &= -({\bf 1}, {\bf {\bar 4}}, {\bf 1}) - \left( ({\bf {\bar 4}}, {\bf 1}, {\bf 1}) + ({\bf 1}, {\bf {4}}, {\bf 2}) + ({\bf{4}}, {\bf 6}, {\bf 1}) \right)x + \cdots, \\
	f_{41}(x) &= -({\bf {4}}, {\bf 1}, {\bf 1}) - \left( ({\bf 1}, {\bf {4}}, {\bf 1}) + ({\bf 6}, {\bf {\bar 4}}, {\bf 1}) + ({\bf {\bar 4}}, {\bf 1}, {\bf 2}) \right)x + \cdots
\end{align}
As we will see explicitly below, finding the $f_{ij}(x)$ up to order $x^n$ is sufficient
to determine the BPS indices up to charge $n \gamma_1$.

\subsection{The bulk BPS indices}

Once the $f_{ij}(x)$ have been determined, 
the next step in determining the spectrum of bulk BPS states with 
phase $\vartheta$ is to construct a generating function $Q_p(x)$ 
attached to each double $S$-wall $p$.
As explained in~\cite{gaiotto2013framed,Gaiotto:2012rg}, 
$Q_p(x)$ is a generating function determining
the jumping behavior of the framed BPS spectrum attached to an interface
between surface defects,
when the phase of the interface crosses $\vartheta$.
It is given by \cite{Gaiotto:2012rg,Hollands:2016kgm} 
\begin{equation} \label{eq:Q-def}
	Q_p(x) = 1 + \tau_p \nu_p.
\end{equation}
In our case there are four double $S$-walls
$p_{13}$, $p_{41}$, $p_{23}$, $p_{42}$,
and thus four functions $Q_{ij}(x)$, 
which are explicitly
\begin{equation} \label{eq:Q-explicit}
Q_{ij}(x) = 1 + x f_{ij}(x) f_{ji}(x).
\end{equation}
The extra factor of $x$ appearing in 
\eqref{eq:Q-explicit} is the product of the basic soliton factors $X_a$ in $\tau$ and $\nu$.

$Q_p(x)$ is the character of a representation of $U(1)\times SU(4)\times SU(4)\times SU(2)$, 
where the $U(1)$ keeps track of the electric charge (exponent of $x$) 
and the rest keeps track of the flavor charge.
This representation is a Fock space built from basic fermionic and bosonic constituents
(fermionic for odd electric charge, bosonic for even charge),
and what we need to do is to extract those constituents.
We define $\bm\alpha_n(p)$ to be the constituent vector space
with electric charge $n$, so that $Q_p(x)$ has an expansion of the form
\begin{equation}
	Q_p(x) = \wedge^*(\bm\alpha_1(p) x) \otimes \text{Sym}^*(-\bm\alpha_2(p) x^2) \otimes \cdots 
\end{equation}
There is a straightforward algorithm to compute the $\bm\alpha_n(p)$ order by order in $n$. The coefficient of $x^1$ in $Q_p(x)$ gives $\bm\alpha_1(p)$. We then formally divide $Q_p(x)$ by the fermionic Fock space generated by $\bm\alpha_1(p) x$. The remaining terms of order $x^2$ give $-\bm\alpha_2(p) x^2$. We then
formally divide out by the bosonic Fock space generated by $-\bm\alpha_2(p) x^2$,
and so on. For example, the expansion of $Q_{41}(x)$ to order $x^2$ is
\begin{align*}
Q_{41}&=1+(\bm{4},\overline{\bm{4}},\bm{1}) x+(2\times(\bm{1},\bm{1},\bm{1}) +(\bm{1},\bm{15},\bm{1})  +(\overline{\bm{4}},\overline{\bm{4}},\bm{2}) + (\bm{6},\bm{10},\bm{1}) \\
&\ \ \  +2\times(\bm{6},\bm{6},\bm{1})+(\overline{\bm{10}},\bm{6},\bm{1}) +(\bm{4},\bm{4},\bm{2}) +(\bm{15},\bm{1},\bm{1}) )x^2+\cdots
 \stepcounter{equation}\tag{\theequation}
\end{align*}
Its expansion  in terms of constituents is
\begin{align*}
Q_{41} &=\wedge^*((\bm{4},\overline{\bm{4}},\bm{1})x) \otimes\text{Sym}^*((2\times(\bm{1},\bm{1},\bm{1}) +(\bm{1},\bm{15},\bm{1})+(\overline{\bm{4}},\overline{\bm{4}},\bm{2})\\
&\ \ \ + 2\times(\bm{6},\bm{6},\bm{1})+(\bm{4},\bm{4},\bm{2})+(\bm{15},\bm{1},\bm{1}) )x^2) \otimes\cdots
 \stepcounter{equation}\tag{\theequation}
\end{align*}
From this expansion we now read off the characters of the constituent vector spaces,
\begin{align}
	\bm{\alpha}_1(p_{41}) &= (\bm{4},\overline{\bm{4}},\bm{1}), \\
	-\bm{\alpha}_2(p_{41}) &= 2\times(\bm{1},\bm{1},\bm{1}) +(\bm{1},\bm{15},\bm{1})+(\overline{\bm{4}},\overline{\bm{4}},\bm{2})+ 2\times(\bm{6},\bm{6},\bm{1})+(\bm{4},\bm{4},\bm{2})+(\bm{15},\bm{1},\bm{1})  .
\end{align}
We record here the answers to first order for all of the double walls, obtained
by expanding the corresponding $Q_{ij}$ to first order in $x$:
\begin{align}
	\bm{\alpha}_1(p_{41}) &= (\bm{4},\overline{\bm{4}},\bm{1}), \\
	\bm{\alpha}_1(p_{32}) &= (\overline{\bm{4}},{\bm{4}},\bm{1}), \\
	\bm{\alpha}_1(p_{13}) &= (\bm{6},{\bm{1}},\bm{2}), \\
	\bm{\alpha}_1(p_{24}) &= (\bm{1},\bm{6},\bm{2}).
\end{align}
Next,
as in \cite{Hollands:2016kgm}, we define
\begin{equation} \label{eq:def-L}
\bm{L}(n\gamma_1) = \sum_{p}\bm{\alpha}_n(p)p_\Sigma.
\end{equation}
The sum in \eqref{eq:def-L} runs over the $4$ 
double $S$-walls $p$, each of which lifts to a chain $p_\Sigma$ on $\Sigma$. 
$\bm{L}(n\gamma_1)$ is a 1-cycle on $\Sigma$ valued in representations of $SU(4)\times SU(4)\times SU(2)$. Its homology class $[\bm{L}(n\gamma_1)]$ is necessarily a multiple of $\gamma_1$.

For example, for $n=1$ we have
\begin{equation}
\sum_p\bm{\alpha}_1(p)=(\bm{1},\bm{6},\bm{2})+(\bm{4},\overline{\bm{4}},\bm{1})+(\bm{6},\bm{1},\bm{2})+ (\overline{\bm{4}},\bm{4},\bm{1}).
\end{equation}
Happily, this is the decomposition of the representation $\bm{56}$ of $E_7$.
For this spectral network, each $p_\Sigma$ is in fact a closed chain, in the class $\gamma_1$,
so \eqref{eq:def-L} becomes
\begin{equation}
	[\bm{L}(\gamma_1)] = \bm{56} [\gamma_1].
\end{equation}
Finally, as in 
\cite{Gaiotto:2012rg,Hollands:2016kgm}, the BPS index is computed as the ratio:
\begin{equation}
\bm{\Omega}(n\gamma) = [\mathbf{L}(n\gamma)]/(n\gamma).
\end{equation}
Thus we find
\begin{equation}
	\bm{\Omega}(\gamma_1) = \bm{56},
\end{equation}
and by similar computations we can compute $\bm\Omega(n \gamma_1)$ for larger $n$.
The fact that these BPS indices turn out to be characters of representations of $E_7$,
not only $SU(4) \times SU(4) \times SU(2)$, constitutes evidence for the expected enhancement of flavor symmetry in this theory.

\section{Results}
\label{sec:results}

Up to $n = 11$, the multiplicity for each representation in $\bm{\Omega}(n\gamma_1)$ turns out to be a positive integer multiple of $(-1)^{(n+1)}n$ (and up to $n = 200$, the unflavored 
$\Omega(n\gamma_1)$ is a positive integer multiple of $(-1)^{(n+1)}n$).
This continues the pattern observed in \cite{Hollands:2016kgm} for the $E_6$ theory.
It is convenient to define a reduced index by dividing out this common factor:
\begin{equation} \label{eq:def-reduced}
\bm{\Omega}_{\red}(n\gamma)=\frac{\bm{\Omega}(n\gamma)}{(-1)^{(n+1)}n}.
\end{equation}
Our results for $\bm{\Omega}_{\red}(n\gamma_1)$ for $1 \le n \le 11$ 
are shown in \autoref{table:electric-indices}.

\setlength\LTleft{-1.35cm}
\begin{longtable}{|lc|}
 \hline
$n$ & $\mathbf{\Omega}_{\red}(n\gamma_1)$  \\ 
 \hline\hline
 1 & $1\times\bm{56}$  \\ 
 \hline
 2 & $1\times\bm{133}+2\times\bm{1} $ \\
 \hline
 3 & $1\times\bm{912}+2\times\bm{56}$ \\
 \hline
 4 & $1\times\bm{8645}+2\times\bm{1539}+6\times\bm{133}+7\times\bm{1}$ \\
 \hline
 5 & $1\times\bm{86184}+2\times\bm{27664}+6\times\bm{6480}+13\times\bm{912}+23\times\bm{56}$ \\ 
  \hline
 6 & $1\times\bm{573440}+2\times\bm{365750}+1\times\bm{253935}+6\times\bm{152152}+13\times\bm{40755}+29\times\bm{8645}+12\times\bm{7371}$\\
 & $+51\times\bm{1539}+16\times\bm{1463}+93\times\bm{133}+79\times\bm{1}$\\
 \hline
 7 & $3\times\bm{3792096}+1\times\bm{3635840}+6\times\bm{2282280}+13\times\bm{861840}+29\times\bm{362880}+12\times\bm{320112}$\\ & $+78\times\bm{86184}+44\times\bm{51072}+107\times\bm{27664}+256\times\bm{6480}+320\times\bm{912}+448 \times\bm{56}$
 \\
 \hline
 8 & $7\times\bm{24386670}+1\times\bm{19046664}+3\times\bm{18372354}+2\times\bm{13728792}+13\times\bm{11316305}$\\&$+29\times\bm{7142499}+12\times\bm{6619239}+78\times\bm{3424256}+100\times\bm{980343}+28\times\bm{617253}$\\&$+146\times\bm{573440}+235\times\bm{365750}+97\times\bm{253935}+21\times\bm{238602}+537\times\bm{152152}$\\&$+163\times\bm{150822}+852\times\bm{40755}+1205\times\bm{8645}+589\times\bm{7371}+1726\times\bm{1539}$\\&$+745\times\bm{1463}+2272\times\bm{133}+1398\times\bm{1}$\\
 \hline
 9 & $9\times\bm{195102336}+14\times\bm{100677808}+1\times\bm{94057600}+13\times\bm{86184000}+29\times\bm{86141440}$\\&$+78\times\bm{63431424}+2\times\bm{32995248}+146\times\bm{21633248}+84\times\bm{14910896}+228\times\bm{13069056} $\\&$+97\times\bm{9480240}+21\times\bm{9405760}+444\times\bm{4522000}+576\times\bm{3792096}+287\times\bm{3635840}$\\&$+1055\times\bm{2282280}+532\times\bm{885248}+1707\times\bm{861840}+3110\times\bm{362880}+1551\times\bm{320112}$\\&$
 +5082\times\bm{86184}+3662\times\bm{51072} +5985\times\bm{27664}+376\times\bm{24320}+11595\times\bm{6480}$\\&$+11009\times\bm{912}+12397\times\bm{56}$\\
 \hline
 10 & $15\times\bm{785674890}+79\times\bm{715309056}+30\times\bm{688400856}+7\times\bm{622396775}+15\times\bm{602350749}$\\&$+6\times\bm{561632400}+1\times\bm{412778496}+146\times\bm{378224640}+212\times\bm{209868813}+22\times\bm{175061250}$\\&$+97\times\bm{163601438}+569\times\bm{132793375}+287\times\bm{130945815}+56\times\bm{109120648}+962\times\bm{72847026}$\\&$+281\times\bm{48316905}+2114\times\bm{24386670}+1387\times\bm{23969792}+484\times\bm{19046664}+939\times\bm{18372354}$\\&$+673\times\bm{13728792}+3122\times\bm{11316305}+1905\times\bm{7482618}+6053\times\bm{7142499}+962\times\bm{6760390}$\\&$+3135\times\bm{6619239}+33\times\bm{5248750}+12854\times\bm{3424256}+13764\times\bm{980343}+2015\times\bm{915705}$\\&$+4489\times\bm{617253}+14710\times\bm{573440}+20735\times\bm{365750}+8745\times\bm{253935}+2676\times\bm{238602}$\\&$+40812\times\bm{152152}+17974\times\bm{150822}+53180\times\bm{40755}+59182\times\bm{8645}+30938\times\bm{7371}$\\&$+72422\times\bm{1539}+36278\times\bm{1463}+75217\times\bm{133}+34869\times\bm{1}$\\
 \hline
 11 & $81\times\bm{5311735000}+19\times\bm{5256879264}+147\times\bm{3993830400}+29\times\bm{3516307200}$\\&$+288\times\bm{2176761600}+553\times\bm{2120058304}+24\times\bm{2032316000}+13\times\bm{1924722800}$\\&$+185\times\bm{1903725824}+1\times\bm{1714199760}+97\times\bm{1700755056}+799\times\bm{985944960}$\\&$+2021\times\bm{789703992}+484\times\bm{656594400}+932\times\bm{619736832}+666\times\bm{462143232}$\\&$+2802\times\bm{339066000}+855\times\bm{236888960}+4440\times\bm{195102336}+34\times\bm{190466640}$\\&$+4848\times\bm{188972784}+2546\times\bm{179262720}+5517\times\bm{100677808}+826\times\bm{94057600}$\\&$
+5914\times\bm{86184000}+10638\times\bm{86141440}+7772\times\bm{67395888}+24528\times\bm{63431424}$\\&$+829\times\bm{32995248}+36469\times\bm{21633248}+7858\times\bm{17926272}+21827\times\bm{14910896}$\\&$+3211\times\bm{14220360}+49219\times\bm{13069056}+21586\times\bm{9480240}+6833\times\bm{9405760}$\\&$+81797\times\bm{4522000}+75935\times\bm{3792096}+42709\times\bm{3635840}+126875\times\bm{2282280} $\\&$+8651\times\bm{2273920}+79317\times\bm{885248}+174663\times\bm{861840}+274960\times\bm{362880}$\\&$+147738\times\bm{320112}+345964\times\bm{86184}+271269\times\bm{51072}+367084\times\bm{27664}$\\&$+42715\times\bm{24320}+601036\times\bm{6480}+475863\times\bm{912}+444394\times\bm{56}
 $\\
 \hline
\caption{\label{table:electric-indices}Reduced indices for charges $n\gamma_1$ in the $E_7$ Minahan-Nemeschansky theory, with flavor information included.}
\end{longtable}

We make a few comments about these results:

\begin{itemize}
\item The result can be compared with~\cite{Huang:2013yta}, which gives refined BPS states of a 5d theory with $E_7$ flavor symmetry obtained by compactifying M-theory on a CY manifold which is a bundle over a del Pezzo surface. It is natural to suspect that further compactifying this theory on $S^1$ to four dimensions would give the Minahan-Nemeschansky $E_7$ theory. As far as the BPS
states go, the precise relation between the 5d and 4d theories 
is not clear; but following \cite{Hollands:2016kgm}
we can find a surprisingly close match by the following ad hoc procedure. We sum the 5d results over spins $j_L$ and $j_R$,
i.e. we just count the total number of multiplets, and compare that with the reduced
index in 4d. For charges $\gamma_1$ and $3\gamma_1$, the 4d and 5d results match exactly. However, for charge $2\gamma_1$ the 4d result contains one more $\bm{1}$ than the 5d, and for $4\gamma_1$ the 4d result has an extra $\bm{133}+\bm{1}$. For charge $5\gamma_1$, the results are different by $193536-192568=968$ in size. It would be natural to try to identify this mismatch
as coming from the multiplets $\bm{912}+\bm{56}$, but comparing our result with the $E_7$ decompositions
given in \cite{Huang:2013yta}, it seems that the mismatch is actually worse:
the representations given in \cite{Huang:2013yta} in 5d are not a subset of the representations we have computed in 4d.
For charge $6\gamma_1$, we only looked at the difference in sizes: it is $3455104 - 3451215=3889$. 
It would be very interesting to understand better what the precise relation between the 5d and 4d results should be.

\item If we omit the flavor information, replacing representations by their dimensions, 
then we can actually solve for the generating functions in closed form, assuming
the symmetries $f_{13}(x)=f_{42}(x),\, f_{31}(x)=f_{24}(x)$. Building the
corresponding $Q$ we obtain:
\begin{align*}
Q_{41}=Q_{23}&=1+\frac{(-1+x+\sqrt{1-34x+x^2})^2}{16x}\\
&=(1+x)^{16}(1-x^2)^{-168}(1+x^3)^{2944}(1-x^4)^{-64752}(1+x^5)^{1573248}+\cdots
 \stepcounter{equation}\tag{\theequation}
\end{align*}
\begin{align*}
Q_{31}=Q_{42}&=\frac{5+5x^2-3\sqrt{1-34x+x^2}+x(-26+3\sqrt{1-34x+x^2})}{2(1+x)^2}\\
&=(1+x)^{12}(1-x^2)^{-102}(1+x^3)^{1664}(1-x^4)^{-35472}(1+x^5)^{845952}+\cdots
 \stepcounter{equation}\tag{\theequation}
\end{align*}
The $Q$'s satisfy algebraic equations:
\begin{gather}
(1+x)^2Q_{42}^2-(5-26x+5x^2)Q_{42}+4(1+x)^2=0, \label{eq:Q-alg-1} \\	
4xQ_{41}^2-(1-10x+x^2)Q_{41}+(1+x)^2=0. \label{eq:Q-alg-2}
\end{gather}
The discriminants both vanish at $(17 + 12 \sqrt{2})^{-1}$. Using the technique in~\cite{Mainiero:2016xaj} and~\cite{Hollands:2016kgm}, these algebraic equations determine the
asymptotic behavior of the BPS degeneracies as
\begin{equation}
|\Omega(n\gamma_1)|\sim cn^{-\frac{5}{2}}(17 + 12 \sqrt{2})^{n},
\end{equation}
for some constant $c$. 
Using \eqref{eq:Q-alg-1}-\eqref{eq:Q-alg-2} 
it becomes possible to compute the
$\Omega(n \gamma_1)$ for much larger $n$, e.g. $1 \le n \le 200$,
and compare with these asymptotics; the agreement is very good.

\end{itemize}

\begin{table}[tbp]
\centering
\begin{tabular}{|lc|c|}
 \hline
 $n$ & $\Omega_{\red}(n\gamma_1)$  \\  
  \hline\hline
 1 & 56  \\ 
 \hline
 2 & 135  \\
 \hline
 3 & 1024  \\
 \hline
 4 & 12528 \\
 \hline
 5 & 193536 \\ 
  \hline
 6 & 3455104 \\ 
 \hline
 7 & 68179968 \\ 
 \hline
 8 & 1447549920 \\ 
 \hline
 9 & 32495488000 \\
 \hline
 10 & 762222261888\\
 \hline
 11 & 18524656253952\\
 \hline
\end{tabular}
\caption{\label{table:numerical-indices} Numerical reduced indices for charges $n \gamma_1$ in
the $E_7$ Minahan-Nemeschansky theory.}
\end{table}

\section{BPS states with charge \texorpdfstring{$\gamma_1+\gamma_2$}{gamma1+gamma2}}
\label{sec:states-charge-11}
\begin{figure}[tbp]
\centering 
\includegraphics[width=.65\textwidth]{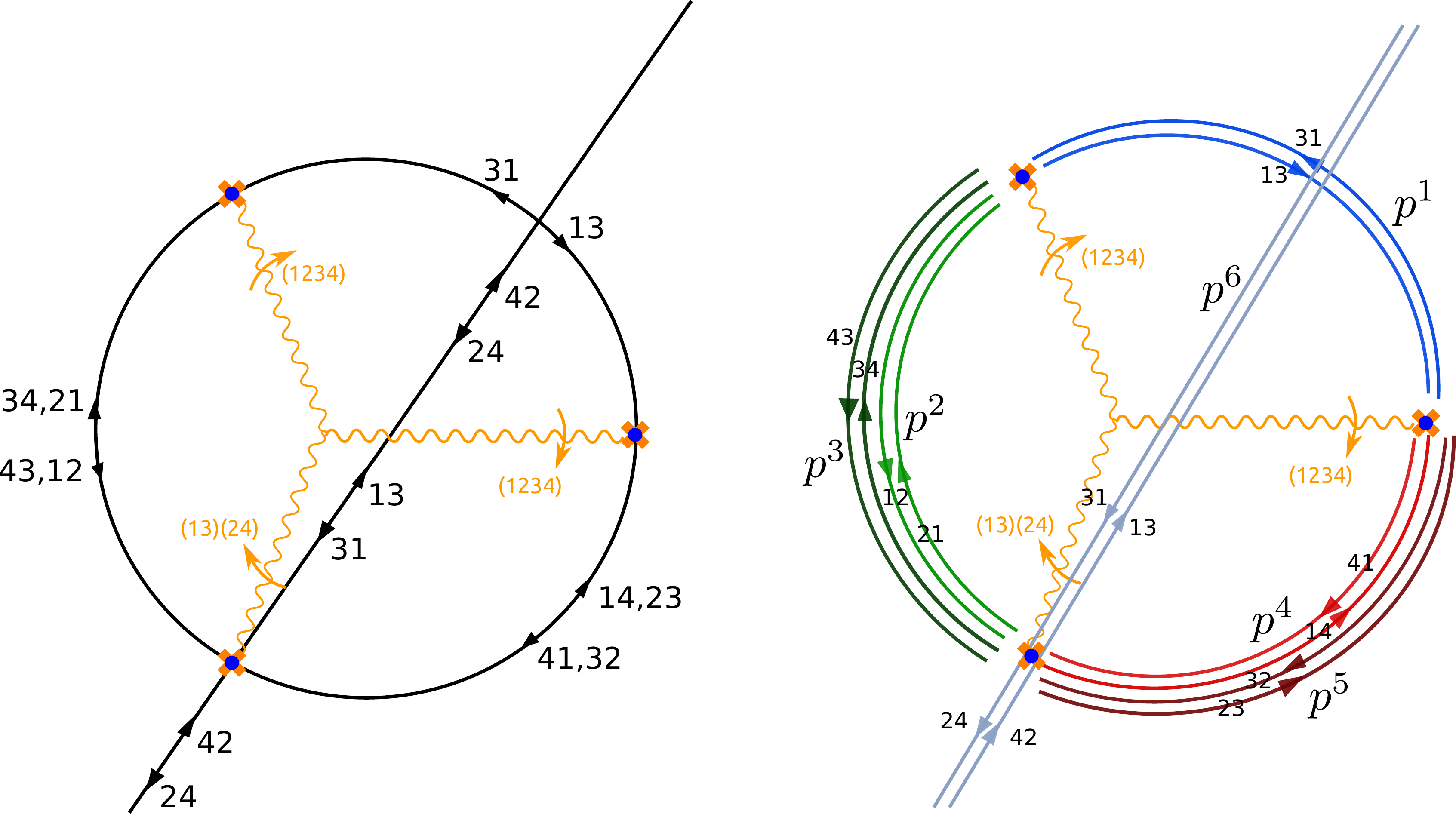}
\caption{\label{fig:charge11} Left: the spectral network of charge $\gamma_1+\gamma_2$. All the walls connecting the same punctures are degenerate. Right: a resolution of the same spectral network.}
\end{figure}

The circular spectral network is the simplest case, but we can use the same method for other charges, at the price of dealing with more complicated spectral networks. 
In this paper we limit ourselves to briefly considering the next simplest case, the BPS states of charge $\gamma_1 + \gamma_2$. The spectral network for charge $\gamma_1 + \gamma_2$ is shown in \autoref{fig:charge11}; it can be computed by the methods we reviewed in
\autoref{sec:building-sn}.

In this case there are six double walls, shown in \autoref{fig:charge11}, 
which we label $p^1, \dots, p^6$.
Carrying out the computations of soliton degeneracies and plethystic logarithms
as in \autoref{sec:BPS states} above, we obtain at leading order
\begin{align}
\bm{\alpha}_{1}(p^1)&=(\bm{6},\bm{6},\bm{1}), \\
\bm{\alpha}_{1}(p^2)&=(\bm{4}\otimes\overline{\bm{4}},\bm{1},\bm{1})+(\bm{4},\bm{4},\bm{2}), \\
\bm{\alpha}_{1}(p^3)&=(\bm{4}\otimes\overline{\bm{4}},\bm{1},\bm{1})+(\overline{\bm{4}},\overline{\bm{4}},\bm{2}), \\
\bm{\alpha}_{1}(p^4)&=(\bm{1},\bm{4}\otimes\overline{\bm{4}},\bm{1})+(\bm{4},\bm{4},\bm{2}), \\
\bm{\alpha}_{1}(p^5)&=(\bm{1},\bm{4}\otimes\overline{\bm{4}},\bm{1})+(\overline{\bm{4}},\overline{\bm{4}},\bm{2}), \\
\bm{\alpha}_{1}(p^6)&=(\bm{1},\bm{1},\bm{2}\otimes\overline{\bm{2}}).
\end{align}
The lifts of paths $p^1$ and $p^6$ are cycles in the class $\gamma_1 + \gamma_2$. However, the lifts of paths $p^2,p^3,p^4$ and $p^5$ do not form closed loops individually. Instead, we get closed cycles with charge $\gamma_1 + \gamma_2$ as combinations of these lifts:
\begin{align}
w^1_\Sigma&=p^2_\Sigma+p^3_\Sigma, \\
w^2_\Sigma&=p^4_\Sigma+p^5_\Sigma, \\
w^3_\Sigma&=p^2_\Sigma+p^4_\Sigma, \\
w^4_\Sigma&=p^3_\Sigma+p^5_\Sigma.
\end{align}
\begin{figure}[tbp]
\centering 
\includegraphics[width=.75\textwidth]{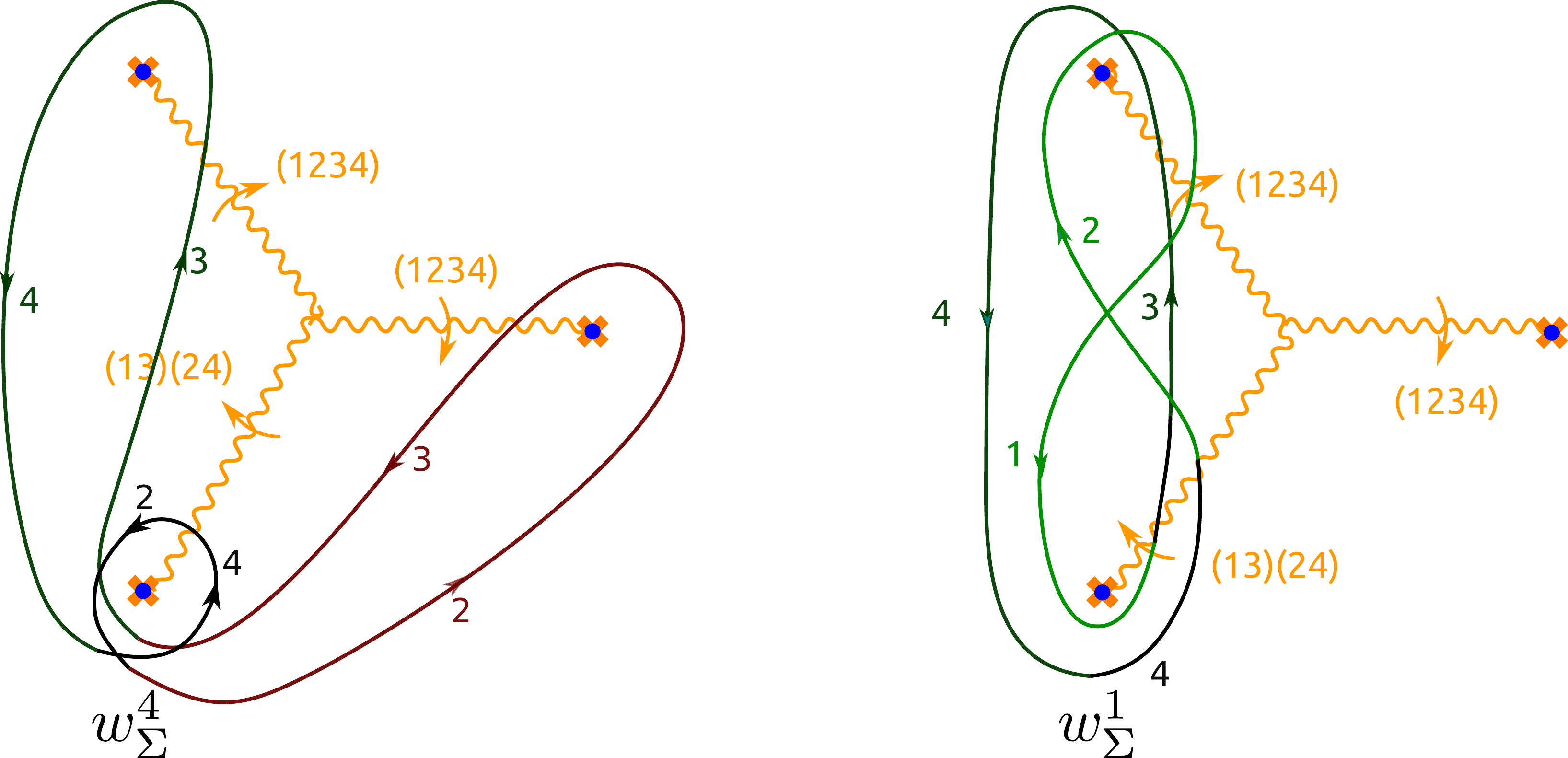}
\caption{\label{fig:sn-11} Building closed cycles with charge $\gamma_1 + \gamma_2$ as
combinations of lifts. Left: $w^4_\Sigma = p^3_\Sigma + p^5_\Sigma$. Right: $w^1_\Sigma = p^2_\Sigma + p^3_\Sigma$.}
\end{figure}
As defined above,
\begin{align*}
\bm{L}(\gamma_1+\gamma_2)&=(\bm{6},\bm{6},\bm{1}){p^1_\Sigma}+(\bm{1},\bm{1},\bm{2}\otimes\overline{\bm{2}}){p^6_\Sigma}+(\bm{4}\otimes\overline{\bm{4}},\bm{1},\bm{1}){w^1_\Sigma}+(\bm{1},\bm{4}\otimes\overline{\bm{4}},\bm{1}){w^2_\Sigma}\\
&\ \ \ +(\bm{4},\bm{4},\bm{2}){w^3_\Sigma}+(\overline{\bm{4}},\overline{\bm{4}},\bm{2}){w^4_{\Sigma}},
 \stepcounter{equation}\tag{\theequation}
\end{align*}
which gives
\begin{align*}
\bm{\Omega}(\gamma_1+\gamma_2)&=(\bm{6},\bm{6},\bm{1})+(\bm{1},\bm{1},\bm{3})+(\bm{15},\bm{1},\bm{1})+(\bm{1},\bm{15},\bm{1})+(\bm{4},\bm{4},\bm{2})\\
&\ \ \ +(\overline{\bm{4}},\overline{\bm{4}},\bm{2})+3\times(\bm{1},\bm{1},\bm{1}).
 \stepcounter{equation}\tag{\theequation}
\end{align*}
This is the decomposition of the representation $\bm{133}+3\times\bm{1}$ of $E_7$, so altogether we have found
\begin{equation}
	\bm\Omega(\gamma_1 + \gamma_2) = {\bm{133} + 3 \times \bm{1}}.
\end{equation}

With computer assistance we calculated the BPS index $\bm\Omega(n(\gamma_{1} + \gamma_2))$,
where $1 \le n \le 5$.
The results are given in \autoref{table:11-indices} below. As before, 
the results are consistent with the string-network 
analysis of~\cite{Distler:2019eky}, and as before, they continue the pattern
of being divisible by $(-1)^{n+1} n$: thus, as before, we give the results for the
reduced BPS index \eqref{eq:def-reduced}.

\setlength\LTleft{-1.35cm}
\begin{longtable}{|lc|}
 \hline
$n$ & $\mathbf{\Omega}_{\red}(n\gamma_1)$  \\ 
 \hline\hline
 1 & $1\times\bm{133}+3\times\bm{1}$  \\ 
 \hline
 2 & $1\times\bm{1539}+2\times\bm{133}+4\times\bm{1} $ \\
 \hline
 3 & $1\times\bm{40755}+2\times\bm{8645}+6\times\bm{1539}+2\times\bm{1463}+11\times\bm{133}+12\times\bm{1}$ \\
 \hline
 4 & $1\times\bm{980343}+2\times\bm{365750}+1\times\bm{253935}+6\times\bm{152152}+2\times\bm{150822}+19\times\bm{40755}$\\&$+27\times\bm{8645}+7\times\bm{7371}+57\times\bm{1539}+27\times\bm{1463}+82\times\bm{133}+67\times\bm{1}$ \\
 \hline
 5 & $1\times\bm{23969792}+3\times\bm{11316305}+2\times\bm{7482618}+6\times\bm{7142499}+19\times\bm{3424256}+42\times\bm{980343}$\\&$+6\times\bm{915705}+12\times\bm{617253}+23\times\bm{573440}+67\times\bm{365750}+29\times\bm{253935}+166\times\bm{152152}$\\&$+78\times\bm{150822}+330\times\bm{40755}+386\times\bm{8645}+149\times\bm{7371}+664\times\bm{1539}+349\times\bm{1463}$\\&$+778\times\bm{133}+498\times\bm{1}$\\
  \hline
\caption{\label{table:11-indices}Reduced indices for charges $n(\gamma_1+\gamma_2)$ in the $E_7$ Minahan-Nemeschansky theory, with flavor information included.}
\end{longtable}

\section{Minahan-Nemeschansky \texorpdfstring{$E_6$}{E6} theory revisited}
\label{sec:MN E6}

As we have mentioned, the Fock space decomposition method for extracting
the $\bm\alpha_n(p)$ from $Q_p$ is a bit more efficient than the method used
in \cite{Hollands:2016kgm}. Thus 
we revisited the $E_6$ theory using the Fock space decomposition method. We were able to compute ${\bf \Omega}(n \gamma_1)$ for $n \le 14$. Our results are presented in \autoref{table:electric-indices-e6} below. For $n \le 7$ the results agree with those in \cite{Hollands:2016kgm};
we include them here just for convenience.

\setlength\LTleft{-1.35cm}
\begin{longtable}{|lc|c|}
 \hline
 $n$ & $\mathbf{\Omega}_{\red}(n\gamma_1)$  \\ 
 \hline\hline
 1 & $\overline{\bm{27}} $ \\ 
 \hline
 2 & $\bm{27}$  \\
 \hline
 3 & $\bm{78}+2\times\bm{1}$  \\
 \hline
 4 & $\overline{\bm{351}}+2\times\overline{\bm{27}}$\\
 \hline
 5 & $\bm{1728}+2\times\bm{351}+6\times\bm{27}$ \\ 
  \hline
 6 & $\bm{5824}+2\times\bm{2925}+\bm{2430}+6\times\bm{650}+13\times\bm{78}+16\times\bm{1}$ \\ 
 \hline
 7 & $\overline{\bm{19305}}+3\times\overline{\bm{17550}}+6\times\overline{\bm{7371}}+13\times\overline{\bm{1728}}+12\times\overline{\bm{351}'}+29\times\overline{\bm{351}}+44\times\overline{\bm{27}}$ \\ 
 \hline
 8 & $1\times\bm{54054}+7\times\bm{51975}+3\times\bm{46332}+2\times\bm{34398}+13\times\bm{17550}+12\times\bm{7722}+29\times\bm{7371}$\\&$+78\times\bm{1728}+28\times{\bm{351}'}+100\times\bm{351}+163\times\bm{27}$\\
 \hline
 9 & $9\times\bm{252252}+1\times\bm{146432}+14\times\bm{105600}+13\times\bm{78975}+29\times\bm{70070}+2\times\bm{43758}$\\&$+78\times\bm{34749}+84 \times\overline{\bm{5824}}+146\times\bm{5824}+21\times\bm{3003}+228\times\bm{2925}+97\times\bm{2430} $\\&$
 +444\times\bm{650}+532\times\bm{78}+376\times\bm{1} $\\
  \hline
 10 & $15\times\overline{\bm{494208}}+7\times\overline{\bm{459459}}+6\times\overline{\bm{412776}}+15\times\overline{\bm{393822}}+30\times\overline{\bm{386100}}+1\times\overline{\bm{359424}'}$\\&$
 +79\times\overline{\bm{314496}}+146\times\overline{\bm{112320}}+22\times\overline{\bm{61425}}+212\times\overline{\bm{51975}}+97\times\overline{\bm{46332}}+56\times\overline{\bm{34398}}$\\&$
 +287\times\overline{\bm{19305}}+569\times\overline{\bm{17550}}+281\times\overline{\bm{7722}}+962\times\overline{\bm{7371}}
 +1387\times\overline{\bm{1728}}+1905\times\overline{\bm{351}}$\\&$+962\times\overline{\bm{351}'}+2015\times\overline{\bm{27}}$\\
 \hline
 11 & $19\times\bm{2088450}+81\times\bm{1640925}+29\times\bm{1253070}+147\times\bm{967680}+1\times\bm{853281}+13\times\bm{741312}$\\&$24\times\bm{579150}+97\times\bm{393822}+185\times\bm{386100}+288\times\bm{359424}+553\times\bm{314496}+799\times\bm{112320}$\\&$+484\times\bm{54054}+2021\times\bm{51975}+932\times\bm{46332}+666\times\bm{34398}+855\times\bm{19305}+34\times\bm{19305}'$\\&$+2802\times\bm{17550}+2546\times\bm{7722}+4848\times\bm{7371}+7772\times\bm{1728}+3211\times\bm{351}'+7858\times\bm{351}$\\&$+8651\times\bm{27}$\\
 \hline
 12&$78\times\bm{5054400}+147\times\bm{4752384}+28\times\bm{3309696}+13\times\bm{3162159}+12\times\bm{3007368}$\\&$+290\times\bm{2977975}+526\times\bm{2453814}+97\times\bm{1911195}+1\times\bm{1896180}+16\times\bm{1559376}$\\&$130\times\bm{1337050}+485\times\bm{972972}+1859\times\bm{852930}+916\times\bm{812175}+650\times\overline{\bm{600600}}$\\&$519\times\bm{600600}+7\times\bm{537966}+36\times\bm{371800}+2270\times\overline{\bm{252252}}+4340\times\bm{252252}$\\&$827\times\bm{146432}+5197\times\bm{105600}+1584\times\bm{85293}+2943\times\overline{\bm{78975}}+5325\times\bm{78975}$\\&$+9433\times\bm{70070}+822\times\bm{43758}+19446\times\bm{34749}+16290\times\overline{\bm{5824}}+21759\times\bm{5824}$\\&$1572\times\overline{\bm{3003}}+4157\times\bm{3003}+28563\times\bm{2925}+12841\times\bm{2430}+43257\times\bm{650}$\\&$+37146\times\bm{78}+17436\times\bm{1}$\\
 \hline
 13 & $288\times\overline{\bm{14017536}}+133\times\overline{\bm{13478400}}+34\times\overline{\bm{12648636}}+469\times\overline{\bm{10378368}}+485\times\overline{\bm{7757100}}$\\&$+49\times\overline{\bm{6675669}}+888\times\overline{\bm{6243237}}+1580\times\overline{\bm{5776056}}+91\times\overline{\bm{5501925}}+623\times\overline{\bm{4582656}}$\\&$+4162\times\overline{\bm{4200768}}+1\times\overline{\bm{4088448}}+40\times\overline{\bm{3281850}}+829\times\overline{\bm{2559843}}+4665\times\overline{\bm{1640925}}$\\&$+1416\times\overline{\bm{1253070}}+4364\times\overline{\bm{1123200}}+7529\times\overline{\bm{967680}}+806\times\overline{\bm{741312}}+10516\times\overline{\bm{494208}}$\\&$5099\times\overline{\bm{459459}}+4462\times\overline{\bm{412776}}+8824\times\overline{\bm{393822}}+16388\times\overline{\bm{386100}}+11674\times\overline{\bm{359424}}$\\&$+1296\times\overline{\bm{359424'}}+37016\times\overline{\bm{314496}}+53396\times\overline{\bm{112320}}+52\times\overline{\bm{100386}}+10794\times\overline{\bm{61425}}$\\&$+8808\times\overline{\bm{54054}}+64712\times\overline{\bm{51975}}+31256\times\overline{\bm{46332}}+20191\times\overline{\bm{34398}}+61577\times\overline{\bm{19305}}$\\&$+105595\times\overline{\bm{17550}}+65758\times\overline{\bm{7722}}+153851\times\overline{\bm{7371}}+174359\times\overline{\bm{1728}}+182898\times\overline{\bm{351}}$\\&$+101325\times\overline{\bm{351'}}+147690\times\overline{\bm{27}}$\\
 \hline
 14 & $258\times\bm{38146680}+472\times\bm{34906950'}+3853\times\bm{30115800}+826\times\bm{26702676}+328\times\bm{22007700}$\\&$827\times\bm{19768320}+566\times\bm{19297278}+1059\times\bm{17918901}+46\times\bm{17453475}+22\times\bm{16992612}$\\&$+21\times\bm{16540524}+68\times\bm{14805504}+3810\times\bm{10378368}+1\times\bm{8401536}+68\times\bm{8281845}$\\&$+9540\times\bm{7601958}+4936\times\bm{7528950}+4300\times\bm{6747300}+8276\times\bm{6243237}+1296\times\bm{6110208}$\\&$+14484\times\bm{5776056}+2780\times\bm{5553900}+778\times\bm{5501925}+4588\times\bm{4582656}+29245\times\bm{4200768}$\\&$+20562\times\bm{2088450}+56\times\bm{1837836}+62331\times\bm{1640925}+24493\times\bm{1253070}+31638\times\bm{1123200}$\\&$+98816\times\bm{967680}+2033\times\bm{853281}+10941\times\bm{741312}+6637\times\bm{638820}+21892\times\bm{579150}$\\&$+36525\times\bm{494208}+18415\times\bm{459459}+14349\times\bm{412776}+57206\times\bm{393822}+103179\times\bm{386100}$\\&$+148984\times\bm{359424}+246082\times\bm{314496}+304906\times\bm{112320}+31152\times\bm{61425}+147492\times\bm{54054}$\\&$490011\times\bm{51975}+239499\times\bm{46332}+174995\times\bm{34398}+253940\times\bm{19305}+14701\times\bm{19305'}$\\&$564835\times\bm{17550}+479896\times\bm{7722}+826668\times\bm{7371}+995856\times\bm{1728}+413309\times\bm{351'}$\\&$+853638\times\bm{351}+703835\times\bm{27}$\\
 \hline
 \caption{\label{table:electric-indices-e6} Reduced BPS indices for the Minahan-Nemeschansky $E_6$ theory.}
\end{longtable}

\appendix

\section{Sign rules} \label{app:signs}

In this appendix we address a tricky question of signs which arises in the 
computation of the BPS indices.

We need to recall a few details from
\cite{Gaiotto:2012rg}. In that paper, the generating function of framed
2d-4d BPS states for a given interface $\wp$ is written as an expansion 
in formal variables, of the form
\begin{equation} \label{eq:F-gen}
	F(\wp) = \sum \underline{\overline\Omega}(\wp, c) X_c.
\end{equation}
Here the index $c$ runs \textit{roughly} over possible charges for a BPS state
of the interface $\wp$, and $\underline{\overline\Omega}(\wp, c)$ 
is \textit{roughly} the BPS index counting states of charge $c$.
However, the precise meaning is a bit subtler, because of ambiguity in defining
the fermion number in a system with only two-dimensional Poincare invariance.
We parameterize our ignorance
by saying $c$ is valued in a $\Z_2$ extension of the naive 
space of charges for the interface, and
letting $H$ denote the generator of the 
extension, we have $\underline{\overline\Omega}(\wp, c + H) 
= -\underline{\overline\Omega}(\wp, c)$.
To compensate this we further define
$X_{c+H} = -X_c$, so that the product $\underline{\overline\Omega}(\wp,c) X_c$
appearing in \eqref{eq:F-gen} is well defined
and independent of how we lift the charge to this extension.

The soliton generating functions $\tau$ and $\nu$ on an $S$-wall
of type $ij$-$ji$ 
are similarly written in terms of formal variables $X_a$, which also
lie in $\Z_2$ extensions of the naive space of soliton charges:
in $\tau$ the extended charges $a$ which appear are charges of solitons from vacuum 
$i$ to vacuum $j$, while in $\nu$ the extended charges $b$ are solitons from vacuum
$j$ to vacuum $i$. 
Given such an $a$ and $b$, there is also a charge
$\mathrm{cl}(a+b)$,
which is an extended 4d charge: it lives in a $\Z_2$ extension of the
lattice of charges for 4d particles. We 
introduce the notation $\mathrm{cl}(X_a X_b) = X_{\mathrm{cl}(a+b)}$.
Then the generating functions
\begin{equation}
Q_p = 1 + \mathrm{cl}(\tau \nu)	
\end{equation}
are functions in the formal variables $X_{\widetilde \gamma}$.
Once we consider purely 4d particles, there is no 
fermion number ambiguity, and thus it is possible
to choose a canonical extended $\widetilde\gamma$ for each ordinary charge $\gamma$.

In \cite{Gaiotto:2012rg} a specific geometric realization of the extended charges is chosen.
The key technical device is to consider paths on the unit tangent bundle 
$\widetilde\Sigma := UT\Sigma$, instead of on $\Sigma$ itself. Then:
\begin{itemize}
	\item All extended charges are homology classes of paths in $\widetilde\Sigma$, considered
	modulo the relation $2H = 0$, where $H$ represents a loop winding once
around a fiber of $\widetilde\Sigma$.
	\item 
The extended charges of states
of an interface $\wp$ are realized as homology classes of paths on 
$\widetilde \Sigma$, ending on the preimages of the tangent vectors to $\wp$ at its endpoints.
\item The extended soliton charges $a$ on a wall of the spectral network 
are realized as homology classes of paths on $\widetilde\Sigma$, 
whose endpoints are tangent vectors
pointing in opposite directions along the wall: the initial vector points in the
direction of decreasing soliton mass, 
while the final vector points in the direction
of increasing mass.
\item The extended 4d charges $\widetilde\gamma$ are realized as homology classes of closed paths
on $\widetilde\Sigma$.
\item A closed path realizing $\mathrm{cl}(a + b)$ is obtained by gluing open paths realizing $a$ and $b$ at their endpoints to make a closed loop on $\widetilde\Sigma$.
\item A canonical lift $\widetilde\gamma$ 
of a homology class $\gamma \in H_1(\Sigma)$ is obtained as follows.
Represent $\gamma$ by an oriented submanifold $P \subset \Sigma$. The oriented 
unit tangent vector field to $P$ gives a lift to
a submanifold $\widetilde P \subset \widetilde\Sigma$. Finally
$\widetilde\gamma = [\widetilde P] + n_P H$ where $n_P$ is the number of
connected components of $P$.
\end{itemize}

Although this realization is canonical and theoretically convenient,
keeping track of lifts to the unit tangent bundle can be annoying, so
it is sometimes useful to switch to an alternate
realization of the extended charges.
In this alternate realization, which we call the
``untwisted formalism,'' instead of $\widetilde\Sigma$ we consider
$\Sigma' = \Sigma \setminus b$, where $b$ is the branch locus of the
covering $\pi: \Sigma \to C$. Then:
\begin{itemize}
\item Extended charges are represented by homology
classes of paths on $\Sigma'$ plus multiples of a formal variable $H$, 
where we impose $2 H = 0$, and also the following
relation: if $L$ is a loop around a branch point with ramification index $n$,
then $L = (n-1)H$.
\item The extended charges of states
of an interface $\wp$ are realized as homology classes of open paths on 
$\Sigma'$, ending on the preimages of the endpoints of $\wp$. 
\item Given a soliton associated to a wall of a spectral network,
its extended charge $a$ is a homology class of paths on $\Sigma'$.
The charge $a$ depends on a choice of a co-orientation of the wall;
if we reverse the co-orientation then the charge $a$ is replaced by $a+H$.
In practice, we generally fix once and for all a co-orientation for each wall.
\item The extended 4d charges $\widetilde\gamma$ are realized as homology classes
of closed paths on $\Sigma'$.
\item A closed path realizing $\mathrm{cl}(a+b)$ is 
obtained by gluing
open paths representing $a$ and $b$. The result of this process is independent
of the co-orientation we choose on the wall, since reversing the co-orientation changes both
$a \to a + H$ and $b \to b + H$, thus changes $\mathrm{cl}(a+b)$ by $2H = 0$.
\item If $\gamma \in H_1(\Sigma)$ obeys $\pi_* \gamma = 0$,
then a canonical lift $\widetilde\gamma$ of $\gamma$ is obtained as follows.
Represent $\gamma$ by an oriented submanifold $P \subset \Sigma'$, such
that $\pi_* P$ has only transverse self-intersections.
Then $\widetilde\gamma = [P] + n'_P H$,
where $n'_P$ is the number of self-intersections of $\pi_* P$.
(For $\gamma$ which do not necessarily obey $\pi_* \gamma = 0$
we cannot get a canonical lift for free, but we can get one after making a choice
of a spin structure on $C$.)
\end{itemize}

The two formalisms are equivalent; however, to construct an explicit 
equivalence between them, one needs to use a spin structure on $C$.

In this paper, as well as in \cite{Hollands:2016kgm}, we work in
the untwisted formalism. We choose once and for all a co-orientation on each
wall. Thus all the formal variables which we use concretely represent
homology classes of paths on $\Sigma'$, and the product of formal variables is
induced from concatenation or addition of homology classes.
For concrete computations we make a choice of a basic soliton charge
$a$ along each wall, and,
as indicated in \eqref{eq:Q-explicit}, we define a formal variable
$x$ by $\mathrm{cl}(X_a X_b) = x$, where $a$ and $b$ are the two solitons
along the wall. After so doing, we have to check carefully
whether $x = X_{\widetilde\gamma}$ or $x = - X_{\widetilde\gamma}$.
According to the rules above, this means we have to draw a loop
representing $a+b$, and count how many self-intersections its projection
to $C$ has: calling this number $k$, we have $x = (-1)^k X_{\widetilde\gamma}$.
In the computations described in 
\autoref{sec:BPS states} of this paper, as well as in 
the main example described in \cite{Hollands:2016kgm},
we chose the basic charges $a$ in such a way that $x = - X_{\widetilde\gamma}$.
This minus sign is ultimately responsible for the fact that when we decompose
$Q_p$ we use fermionic constituents for odd charges and bosonic for even charges;
if $x = X_{\widetilde\gamma}$ instead then we would use bosonic constituents
for all charges.

\bibliography{e7-paper}
\bibliographystyle{utphys}
\end{document}